\documentclass[
  aps,
  prl,
  amsmath,
  reprint,
  superscriptaddress,
  letterpaper,
  floatfix,
  longbibliography
]{revtex4-2}

\usepackage{graphicx,color}
\usepackage{verbatim}
\usepackage{amssymb}
\usepackage{amsmath}
\usepackage{amsfonts}
\usepackage{mathdots}
\usepackage{braket}
\usepackage{bm}
\usepackage{float}
\usepackage{bbm}
\usepackage{soul}

\usepackage{xcolor}
\definecolor{apslinkblue}{RGB}{46,48,146}

\usepackage[
  colorlinks=true,
  linkcolor=apslinkblue,
  citecolor=apslinkblue,
  urlcolor=apslinkblue
]{hyperref}

\DeclareMathOperator{\Tr}{Tr}

\newcommand{\ii}{\mathrm{i}}
\newcommand{\bk}{\mathbf{k}}
\newcommand{\bR}{\mathbf{R}}

\newcommand{\bS}{\mathbf{S}}
\newcommand{\bs}{\mathbf{s}}
\newcommand{\bsigma}{\bm{\sigma}}

\begin{document}

\title{Quantum Geometric Kondo Cloud}

\author{Grant Z. X. Yang} \thanks{Contact author: zyangdi@connect.ust.hk}
\affiliation{Department of Physics, Hong Kong University of Science and Technology, Clear Water Bay, Hong Kong, China}

\author{K. T. Law} \thanks{Contact author: phlaw@ust.hk}
\affiliation{Department of Physics, Hong Kong University of Science and Technology, Clear Water Bay, Hong Kong, China}

\date{\today}

\begin{abstract}
A magnetic impurity embedded in a metal is collectively screened by Fermi-surface quasiparticles into a many-body spin-singlet ground state, forming a Kondo cloud of size $\xi_{\rm K}\sim\hbar v_F/(k_B T_{\rm K})$.
This kinematic picture collapses in flat bands, where $v_F=0$ and the hierarchy of dispersive energy shells is absent.
Here we show that the missing organizing principle is quantum geometry.
A magnetic impurity coupled to an isolated flat band selects a single active bath mode: a coherent superposition of flat-band Bloch states weighted by the hybridization factor $v(\mathbf{k})$, while all orthogonal flat-band modes remain dark.
The resulting flat-band Kondo problem is a quantum geometric molecule, with an algebraic Kondo scale set by the total projected hybridization strength rather than a logarithmic-renormalization scale.
In real space, the impurity-bath spin correlation defines a quantum geometric Kondo cloud.
Its cloud-size tensor admits a gauge-invariant decomposition into a hybridization-weighted quantum metric, a dressed Berry-connection covariance, and a positive hybridization-gradient term, yielding the lower bound
$\xi_{\rm K}^2\geq \sum_{\mathbf{k}}\rho(\mathbf{k}){\rm Tr}\,g(\mathbf{k})$,
where $\rho(\mathbf{k})=|v(\mathbf{k})|^2/\sum_{\mathbf{k}}|v(\mathbf{k})|^2$.
Our result reveals that, in flat bands, Kondo screening is governed by the quantum geometry and interference structure of the impurity-selected Bloch wave packet, rather than Fermi-surface kinematics.
\end{abstract}

\maketitle

\emph{Introduction.}\textemdash
A magnetic impurity embedded in a metal is screened by surrounding bath electrons into a many-body spin singlet, forming the spatially extended Kondo cloud~\cite{kondo1964resistance,sorensen1996scaling,anderson1961localized,barnes1976new,hewson1997kondo,gubernatis1987spin,affleck2001detecting,sorensen2005kondo,hand2006spin,affleck2008friedel,pereira2008kondo,barzykin1996kondo,borda2007kondo,v2020observation}. In the conventional metallic bath, the screening process is controlled by logarithmic renormalization-group (RG) flow near the Fermi surface: the exchange coupling grows under the reduction of the electronic bandwidth and leads to the exponentially small Kondo temperature $T_{\text{K}}$~\cite{anderson1970poor}, giving rise to the screening length as $\hbar v_F/k_B T_{\text{K}}$,
where $v_F$ is the Fermi velocity~\cite{sorensen1996scaling,barzykin1996kondo,borda2007kondo,v2020observation,gubernatis1987spin,affleck2001detecting,sorensen2005kondo,hand2006spin,affleck2008friedel,pereira2008kondo}. This relation makes the central role of Fermi-surface kinematics explicit.
Experimentally, the spatial nature of Kondo screening has been revealed from local Fano--Kondo resonances in magnetic adatoms and quantum corrals~\cite{manoharan2000quantum,knorr2002kondo} to more direct evidence of a mesoscopic screening cloud in quantum-dot interferometers~\cite{v2020observation}. 
Recent atomically resolved measurements in monolayer $\mathrm{MoS_2}$ further showed that Kondo screening can be spatially modulated by the host electronic structure~\cite{van2024modulated}. 
Meanwhile, gate-tunable heavy fermions and Kondo breakdown in moir\'e Kondo lattices have brought Kondo correlations into flat-band platforms~\cite{zhao2023gate,zhao2024emergence}. 
Together, these experiments motivate a spatial perspective on Kondo screening in flat bands.

Moreover, recent advances in moir\'e materials with experimental observations of exotic correlated phases have sparked intense interest in flat-band systems~\cite{cao2018unconventional,cao2018correlated}, challenging the conventional understanding of many-body physics in the absence of Fermi-surface kinematics~\cite{bistritzer2011moire,po2018origin,song2022magic}. In such systems, a (nearly) flat band is isolated from other bands with a band gap $\Delta_{\text{gap}}$, while the interaction strength is smaller than band gap but larger than the flat-band width. The flat-band ordering instabilities can be organized by quantum-geometric nesting~\cite{han2024quantum,zhang2026identifying,sun2025flat}, as a counterpart of Fermi-surface nesting. Most notably, the quantum metric of flat-band superconductors controls both the superfluid weight and coherence length of Cooper pairs~\cite{hu2019geometric,peotta2015superfluidity,chen2024ginzburg,gao2026bootstrapping,liang2017band,herzog2022superfluid,huhtinen2022revisiting,hofmann2023superconductivity,hu2025anomalous}. Furthermore, the quantum metric defines an intrinsic length scale that governs Josephson effects~\cite{li2025flat} and disordered transport in flat bands~\cite{dai2026quantummetriclocalizationquantum,chau2026quantum}, replacing Fermi-surface kinematics.

\begin{figure}[h]
    \centering
    \includegraphics[scale=0.63]{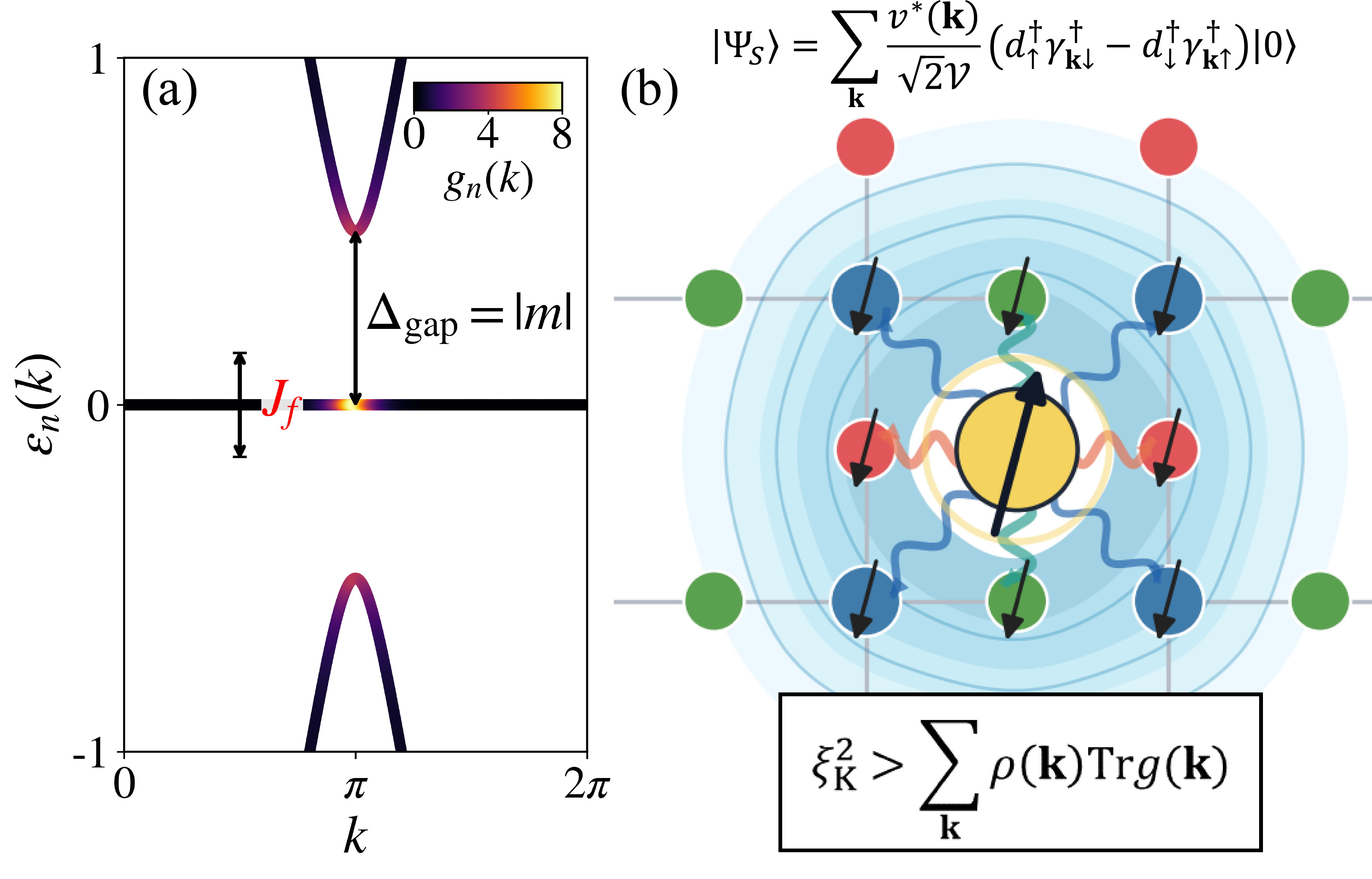}
    \caption{(a) Band dispersions $\varepsilon_n(k)$ of the one-dimensional Lieb lattice at $m/t=0.5$, with color indicating the band quantum metric
    $g_n(k)$. The flat band is isolated from dispersive bands with a large band gap $\Delta_{\text{gap}}\gg J_f$. 
    (b) Schematic illustration of the quantum geometric Kondo cloud.
    The magnetic impurity couples to an active bath mode \(f\), a coherent superposition of flat-band Bloch states weighted by \(v(\bk)\).
    The spatial extent of the Kondo cloud is lower bounded by the hybridization-weighted quantum metric.}
    \label{fig:1}
\end{figure}

Flat bands eliminate the energetic hierarchy of dispersive Bloch states while retaining a nontrivial hierarchy in wave-function geometry. Thus, in the absence of energy-shell renormalization, Kondo screening can no longer be organized by Fermi-surface kinematics. While previous numerical work identified molecular Kondo singlets in flat bands~\cite{tran2018molecular}, the unresolved problem is not merely whether a singlet forms, but what replaces the conventional Kondo cloud as a spatial many-body object.

In this Letter, we show that the replacement is a quantum geometric Kondo cloud.
Starting from a multi-orbital Anderson impurity model, we perform a Schrieffer-Wolff (SW) transformation and project onto an isolated flat band, which factorizes the exchange vertex.
The impurity therefore couples only to an active bath mode, a coherent superposition of flat-band Bloch states weighted by the hybridization factor, while all orthogonal modes remain dark. The molecular Kondo temperature $T_{\mathrm K}$ is set by the singlet-binding energy rather than a logarithmically generated RG scale. The active bath mode exhibits a nontrivial Wannier-like structure in real space, capturing both the impurity-bath interference and the intrinsic spread of flat-band Bloch states. Its ground-state spin correlations establish the quantum geometric Kondo cloud. We show that the cloud-size tensor admits a gauge-invariant decomposition and is bounded from below by the hybridization-weighted quantum metric, independent of the exchange strength. Applying the formalism to a one-dimensional Lieb lattice, we demonstrate that orbital-selective hybridization coherently tunes both the Kondo scale and cloud size, identifying quantum geometry and impurity-bath interference, rather than Fermi-surface kinematics, as the organizing principle of Kondo screening in flat bands.

\emph{Multi-orbital Anderson impurity model.}\textemdash
We start from a spin-$1/2$ Anderson impurity coupled to a multi-orbital bath,
\begin{equation}
\begin{split}
    H_{\mathrm{AIM}}
    &=
    H_d+H_b+H_V,\\
    H_d
    &=
    \varepsilon_d\sum_{\sigma}d^\dagger_\sigma d_\sigma
    +
    U n_{d\uparrow}n_{d\downarrow},\\
    H_b
    &=
    \sum_{\bk\alpha\beta\sigma}
    c^\dagger_{\bk\alpha\sigma}
    \left[
    h_{\alpha\beta}(\bk)-\mu\delta_{\alpha\beta}
    \right]
    c_{\bk\beta\sigma},\\
    H_V
    &=
    \sum_{i\alpha\sigma}
    \left(
    V_{i\alpha}d^\dagger_\sigma c_{i\alpha\sigma}
    +
    V^*_{i\alpha}c^\dagger_{i\alpha\sigma}d_\sigma
    \right).
\end{split}
\label{eq:AIM}
\end{equation}
Here $d_\sigma$ annihilates an impurity electron and $\sigma$ is the spin index, $U>0$ is the on-site repulsion, and $\varepsilon_d<0$ is the impurity level. $c_{\bk \alpha \sigma}$ annihilates a bath electron in the orbital $\alpha$, the bath Hamiltonian can be diagonalized as $H_b=\sum_{\bk n\sigma}\xi_n(\bk)\gamma^\dagger_{\bk n\sigma}\gamma_{\bk n\sigma}$ by a unitary transformation $c_{\bk\alpha\sigma}=\sum_n u_{\alpha n}(\bk)\gamma_{\bk n\sigma}$ with the band index $n$, and $\xi_n(\bk)=\varepsilon_n(\bk)-\mu$ is measured from the chemical potential. Using Fourier transformation $
    c_{i\alpha\sigma}
    =
    \frac{1}{\sqrt{N}}
    \sum_{\bk}
    e^{\ii\bk\cdot\bR_i}
    c_{\bk\alpha\sigma}$,
where $N$ is the number of unit cells, the hybridization Hamiltonian becomes
\begin{equation}
    H_V
    =
    \sum_{\bk n\sigma}
    \left[
    v_n(\bk)d^\dagger_\sigma\gamma_{\bk n\sigma}
    +
    v^*_n(\bk)\gamma^\dagger_{\bk n\sigma}d_\sigma
    \right],
\label{eq:HVband}
\end{equation}
where $v_n(\bk)=\frac{1}{\sqrt{N}}\sum_{\alpha,i}e^{i\bk\cdot\bR_i}V_{i\alpha} u_{\alpha n}(\bk)$ is the band-resolved hybridization factor.
For a single impurity at $\bR_i=0$, $v_n(\bk)\propto\sum_\alpha V_\alpha u_{\alpha n}(\bk)$. Thus, even a local impurity acquires a nontrivial momentum-dependent form factor, describing interference between bath Bloch states and the magnetic impurity.

In the local-moment regime, charge fluctuations of the impurity are virtual. We assume 
$E_0=-\varepsilon_d>0$ and $E_2=\varepsilon_d+U>0$ 
are the excitation energies from the singly occupied impurity to the empty and doubly occupied impurity charge states, respectively. A SW transformation integrating out the high-energy states of impurity gives the effective Kondo Hamiltonian~\cite{supple}
\begin{equation}
    H_{\text{K}}
    =
    \sum_{\bk n\sigma}
    \xi_n(\bk)
    \gamma^\dagger_{\bk n\sigma}\gamma_{\bk n\sigma}
    +
    \sum_{\bk n,\bk' n'}
    J_{nn'}(\bk,\bk')
    \bS_d\cdot \bs_{nn'}(\bk,\bk'),
\label{eq:HK_general}
\end{equation}
where the bath spin operator is defined as $\bs_{nn'}(\bk,\bk')=\frac12 \sum_{\sigma\sigma'}\gamma^\dagger_{\bk n\sigma}\bsigma_{\sigma\sigma'}\gamma_{\bk'n'\sigma'}$ with Pauli matrix $\bsigma=(\sigma_x, \sigma_y, \sigma_z)$ and, for $|\xi_{n}(\bk)|\ll E_{0,2}$, the exchange vertex is 
$J_{nn'}(\bk,\bk')\simeq 2v_n^*(\bk)v_{n'}(\bk')\left(E_0^{-1}+E_2^{-1}\right)$.
The potential-scattering term is omitted since it vanishes at impurity particle-hole symmetry and does not affect the flat-band spin singlet.

\emph{Flat-band projection and exact molecular Kondo solution.}\textemdash
We set the chemical potential $\mu$ at the target isolated flat band as $\xi_0(\bk)=0$ for all $\bk$, denoted by $n=0$. The band gap between the flat band and other dispersive bands is $\Delta_{\text{gap}} = \min_{\bk, n\ne 0}|\xi_n(\bk)|$. We assume the energy hierarchy: $J_f \ll \Delta_{\text{gap}} \ll E_0, E_2$ as shown in Fig.~\ref{fig:1}(a), where $J_f$ is the effective exchange and singlet--triplet spin-gap scale. We now project Eq.~\eqref{eq:HK_general} to a single isolated flat band and neglect the band index for notational simplicity as $\ket{u_{\bk0}}\rightarrow \ket{u_{\bk}}$. The Kondo exchange vertex factorizes as $J(\bk,\bk')=2v^*(\bk)v(\bk')\left(\frac{1}{E_0}+\frac{1}{E_2}\right)$.
We define the total projected hybridization strength
\begin{equation}
    \mathcal{V}^2
    =
    \sum_{\bk}|v(\bk)|^2
    =
    \sum_{\bk\alpha\beta}
    v_\alpha(\bk)
    \mathcal{P}_{\alpha\beta}(\bk)
    v^*_\beta(\bk),
\label{eq:V2}
\end{equation}
where $\mathcal{P}_{\alpha\beta}(\bk) = \left<{u_{\bk \alpha}}|{u_{\bk}}\right>\left<{u_{\bk}}|{u_{\bk\beta}}\right>$ is the flat-band projector. Equivalently, $\mathcal{V}^2$ can be interpreted as an integrated hybridization spectral weight of the flat-band impurity self-energy, $-\text{Im}\Sigma^R(\omega) = \pi \mathcal{V}^2 \delta(\omega)$~\cite{supple,bulla2008numerical}. We then introduce the normalized active bath fermion as a coherent superposition of flat-band Bloch states
\begin{equation}
    f_\sigma
    =
    \frac{1}{\mathcal{V}}
    \sum_{\bk}
    v(\bk)\gamma_{\bk\sigma},
    \qquad
    f^\dagger_\sigma
    =
    \frac{1}{\mathcal{V}}
    \sum_{\bk}
    v^*(\bk)\gamma^\dagger_{\bk\sigma},
\label{eq:f_operator}
\end{equation}
which satisfies $\{f_\sigma,f^\dagger_{\sigma'}\}=\delta_{\sigma\sigma'}$. 
The effective Kondo model in Eq.~\eqref{eq:HK_general} reduces exactly to
\begin{equation}
    H^{\mathrm{flat}}_{\text{K}}
    =
    J_f \bS_d\cdot \bs_f.
\label{eq:Hflat}
\end{equation}
where $\bs_f=\frac{1}{2}\sum_{\sigma\sigma'}f^\dagger_\sigma\bsigma_{\sigma\sigma'}f_{\sigma'}$ is the spin operator for the active bath mode and positive effective exchange coupling $ J_f=2\mathcal{V}^2\left(\frac{1}{E_0}+\frac{1}{E_2}\right) > 0$, ensuring the antiferromagnetic exchange. At impurity particle-hole symmetry, $\varepsilon_d=-U/2$, this becomes $J_f=8\mathcal{V}^2/U$. 
The flat-band operators orthogonal to $f_\sigma$ are dark modes: they carry the remaining macroscopic flat-band degeneracy and determine the thermodynamic filling, but do not couple to the impurity.
Meanwhile, the active-mode occupation $n_f=\sum_\sigma f^\dagger_\sigma f_\sigma$
is conserved. 
The sectors \(n_f=0,2\) are spinless, hence only the \(n_f=1\) sector supports Kondo screening, corresponding to single occupancy of a normalized collective mode of flat-band Bloch states.
Indeed, using Eq.~\eqref{eq:f_operator}, the singlet ground state is
\begin{equation}
\begin{split}
    \ket{\Psi_S}
    &=
    \frac{1}{\sqrt{2}}
    \left(
    d^\dagger_\uparrow f^\dagger_\downarrow
    -
    d^\dagger_\downarrow f^\dagger_\uparrow
    \right)
    \ket{0} \\
    &=
    \sum_{\bk}
    \frac{v^*(\bk)}{\sqrt{2}\mathcal{V}}
    \left(
    d^\dagger_\uparrow \gamma^\dagger_{\bk\downarrow}
    -
    d^\dagger_\downarrow \gamma^\dagger_{\bk\uparrow}
    \right)
    \ket{0}.
\end{split}
\label{eq:singlet_f}
\end{equation}
Equation~\eqref{eq:singlet_f} shows that the ground state is a coherent superposition of singlet states over the whole Brillouin zone (BZ) weighted by a coefficient $v^*(\bk)/\mathcal{V}$, while $\rho(\bk)=|v(\bk)|^2/\mathcal{V}^2$ can be introduced as the momentum-dependent hybridization distribution. Hence, the impurity is screened by a single collective bath mode whose phase coherence is determined by the interference between impurity and bath Bloch states.

The singlet energy of Eq.~\eqref{eq:Hflat} is
$E_S=-3J_f/4$, whereas the triplet manifold has energy
$E_T=J_f/4$. The singlet--triplet spin gap is therefore
$\Delta_{\mathrm{ST}}\equiv E_T-E_S=J_f$.
At impurity particle--hole symmetry, the spinless active-mode sectors $n_f=0,2$ have zero energy. We therefore identify the flat-band molecular Kondo temperature, as the energy required to break the molecular singlet into the lowest unscreened active-charge sector
\begin{equation}
    k_B T_{\mathrm K}^{\mathrm{flat}}
    \equiv
    \Delta_{\mathrm{mol}}
    =
    \frac{3}{4}J_f
    \propto\mathcal V^2.
\label{eq:TKflat}
\end{equation}
Thus the molecular Kondo scale is algebraic in the total projected
hybridization strength. 
At impurity particle--hole symmetry, this becomes
$k_BT_{\mathrm K}^{\mathrm{flat}}=6\mathcal V^2/U$.
The total projected hybridization is an interference quantity: it depends on both local coupling amplitude $|V_{i\alpha}|$ and its coherence with flat-band Bloch wave functions through $v(\bk)$. The flat-band Kondo scale is therefore tunable by both the magnitude and relative phase of the impurity-bath couplings. As shown below, destructive impurity-bath interference can enforce $v(\bk)=0$, rendering a dark channel with no active screening mode and zero Kondo scale, despite finite local hybridization on the lattice sites.

The coherence of the active mode is the microscopic origin of the spatial Kondo cloud in a flat band, despite the absence of a Fermi velocity. Fourier transforming the ground-state wavefunction in Eq.~\eqref{eq:singlet_f} yields a Wannier-like real-space structure $\Phi_\alpha(\mathbf R_i)$, whose spatial profile is fixed by the coherent interference of $v(\bk)u^*_\alpha(\bk)$ over the BZ.
The resulting spin-correlation profile is therefore not a Fermi-surface screening envelope, but the real-space image of an impurity-selected Bloch superposition.
Consequently, the flat-band Kondo cloud cannot, in general, be made arbitrarily short ranged,
its minimal spread is instead bounded by the $v(\bk)$-weighted quantum geometry of the target band, as demonstrated below.

\emph{Quantum geometric Kondo cloud.}\textemdash
The Kondo cloud is the real-space structure of the many-body screening state, where the impurity-bath spin correlation is the standard diagnostic~\cite{gubernatis1987spin,borda2007kondo,barzykin1996kondo,sorensen1996scaling,affleck2001detecting,sorensen2005kondo,hand2006spin,affleck2008friedel,pereira2008kondo}. In a metallic bath, itinerant electrons give the familiar coherence length $\hbar v_F/k_B T_{\text{K}}$, reflecting the velocity scale $v_F$ and the characteristic energy scale $T_{\mathrm K}$. To investigate the real-space profile of the quantum geometric Kondo cloud, we rewrite the active bath mode of flat band in real space as $f_\sigma = \sum_{i\alpha}\Phi_\alpha(\bR_i)c_{i\alpha\sigma}$ with
\begin{equation}
\begin{split}
    &\Phi_{\alpha}(\mathbf{R}_i) = \frac{1}{\mathcal{V}\sqrt{N}}\sum_{\mathbf{k}} e^{-i\mathbf{k\cdot R}_i} v(\mathbf{k}) u^*_{\alpha}(\mathbf{k}).
\end{split}
\label{eq:Phi}
\end{equation}
Replacing $v(\bk)/\mathcal{V}\rightarrow e^{i\bk\cdot\bR_j}/\sqrt{N}$ gives rise to the usual Wannier function $W_{n\alpha}(\bR_j-\bR_i)$ of the target band in orbital basis. This Wannier-like structure shows that the impurity couples to a delocalized superposition of impurity-selected localized orbitals.

The impurity-bath spin correlation is given by $
    C_{i\alpha}
    =
    \braket{\bS_d\cdot\bs_{i\alpha}}
    -
    \braket{\bS_d}\cdot\braket{\bs_{i\alpha}}$
with local bath spin density as $\bs_{i\alpha}=\frac{1}{2}
    c^\dagger_{i\alpha\sigma}
    \bsigma_{\sigma\sigma'}
    c_{i\alpha\sigma'}$.
For the coherent singlet ground state $\ket{\Psi_S}$, both the impurity and bath spin polarizations vanish, $\braket{\bS_d}=\braket{\bs_{i\alpha}}=0$. 
The impurity-bath correlation is thereby~\cite{supple}
\begin{equation}
    C_{i\alpha}=-\frac{3}{4}|\Phi_\alpha(\bR_i)|^2<0.
    \label{eq:spin-correlation}
\end{equation}
Equation~\eqref{eq:spin-correlation} indicates the Kondo-cloud profile is identical to the real-space probability distribution of the active bath mode, with the negative sign encoding the antiferromagnetic nature of the Kondo singlet.

We next demonstrate the real-space profile of the Kondo cloud explicitly. The cloud center is $
    \bR_K
    =
    \sum_{i\alpha}
    |\Phi_\alpha(\bR_i)|^2\bR_i$,
and the cloud-size tensor is defined as
\begin{equation}
    \xi^2_{K,\mu\nu}
    =
    \sum_{i\alpha}
    |\Phi_\alpha(\bR_i)|^2
    (R_i^\mu-R_K^\mu)
    (R_i^\nu-R_K^\nu).
\label{eq:size_tensor_real}
\end{equation}
Here $\mu,\nu$ are the spatial indices. We adopt the orbital-center convention, where the orbital embeddings are measured from the unit-cell center. A general orbital embedding is recovered by replacing $\bR_i\rightarrow \bR_i+\bm{\tau}_\alpha$. We emphasize that quantum geometric tensor and other geometry-dependent quantities are orbital embedding dependent~\cite{simon2020contrasting}. The scalar cloud size is given by the trace of cloud-size tensor as $\xi_{\text{K}}=\sqrt{\Tr\,\xi^2_{\text{K},\mu\nu}}$.

\begin{figure}[t]
    \hspace{-0.5cm}
    \includegraphics[scale=0.75]{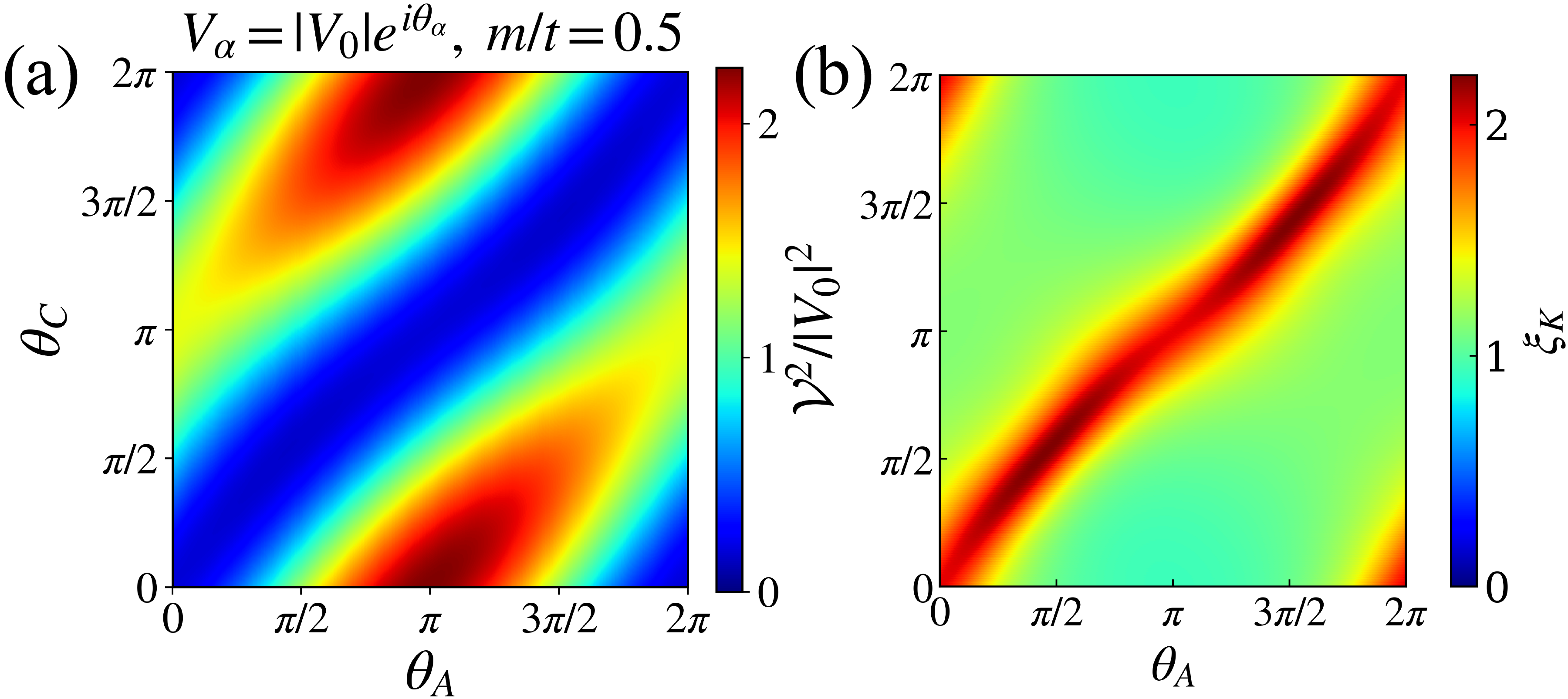}
    \caption{Phase control of the screening energy and length. (a) Projected hybridization strength $\mathcal{V}^2/|V_0|^2$ and (b) Kondo-cloud length $\xi_{\text{K}}$ for equal-amplitude impurity couplings $V_\alpha=|V_0|e^{i\theta_\alpha}$ at $m/t=0.5$. The redundant global hybridization phase is fixed by $\theta_B=0$. Both quantities depend only on the remaining relative phases. The projected hybridization is maximal for
    $\theta_C-\theta_A=\pi$ and $\theta_C=\theta_B$ modulo $2\pi$,
    demonstrating coherent control of the screening gap and spatial cloud through orbital interference.}
    \label{fig:2}
\end{figure}

\begin{figure*}[t]
    \centering
    \includegraphics[scale=0.85]{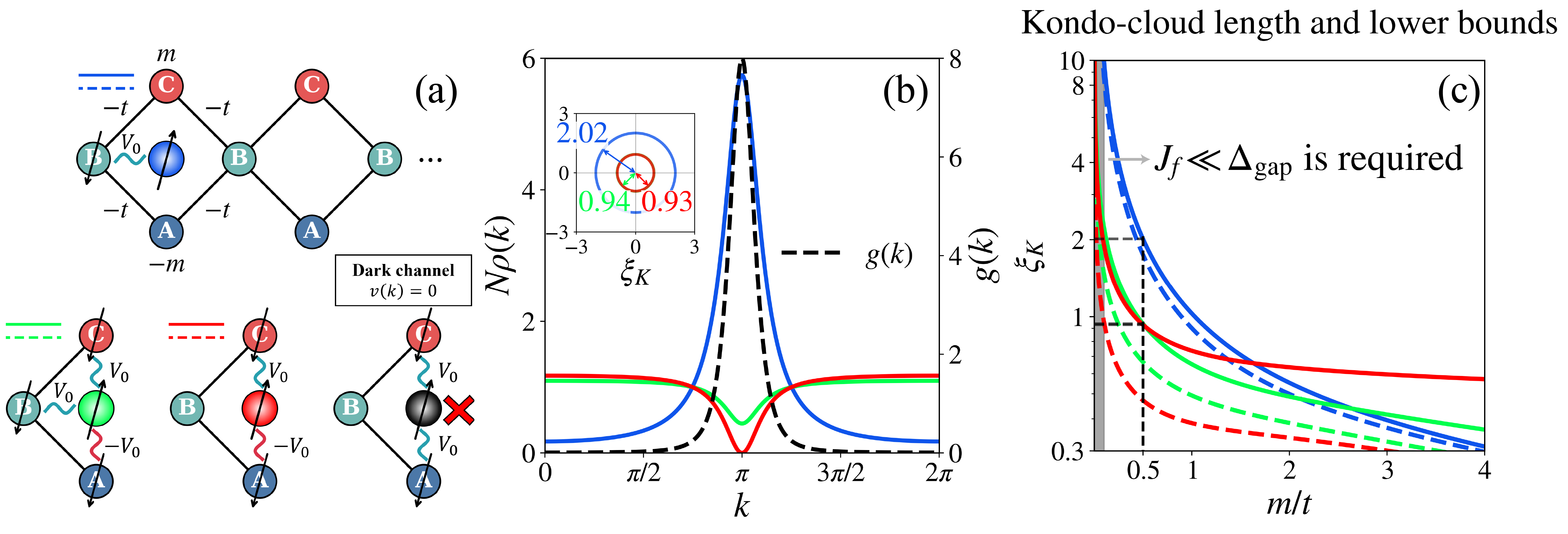}
\caption{
Orbital-selective quantum-geometric Kondo cloud.
(a) One-dimensional Lieb lattice and four impurity channels:
$V_B=V_0$ (blue);
$V_B=V_C=V_0$, $V_A=-V_0$ (green);
$V_C=-V_A=V_0$, $V_B=0$ (red); and
$V_A=V_C=V_0$, $V_B=0$ (black).
The last choice is dark because its projected form factor vanishes
identically.
(b) Normalized distributions $N\rho(k)$ of the three non-dark
channels (left axis) and flat-band metric $g_0(k)$ (black dashed,
right axis) at $m/t=0.5$.
The inset schematically compares the corresponding cloud radii
$\xi_{\rm K}/a=2.02$, $0.94$, and $0.93$.
(c) Cloud length $\xi_{\rm K}/a$ (solid) and metric lower bound
$\xi_{{\rm K},{\rm lb}}/a$ (dashed) versus $m/t$, on a logarithmic
vertical scale. The vertical dashed line marks $m/t=0.5$ used in
panel (b). The gray region lies outside the criterion $J_f\ll \Delta_{\rm gap}$.}
\label{fig:3}
\end{figure*}
Let ${v(\bk)}/\mathcal{V}=\sqrt{\rho(\bk)}e^{i\chi(\bk)}$
with Berry connection $\mathcal{A}_\mu(\bk)=i\braket{u_{\bk}|\partial_\mu u_{\bk}}$,
the dressed Berry connection is defined as 
$X_\mu(\bk)=\mathcal{A}_\mu(\bk)+\partial_\mu\chi(\bk)$.
If $v(\bk)$ has isolated nodes, $\chi(\bk)$ and $X_\mu(\bk)$ are
understood patchwise away from the nodal set.
Here the dressed Berry connection $X_{\mu}(\bk)$ is invariant under the $U(1)$ gauge transformation of the flat-band Bloch state: $\ket{u_{\bk}}\rightarrow e^{i\phi(\bk)}\ket{u_{\bk}}$, and describes the phase information of both Bloch states and hybridization factor. Following Parseval's theorem, the cloud-size tensor yields a gauge-invariant decomposition as~\cite{supple}
\begin{equation}
\begin{split}
    \xi^2_{K,\mu\nu}
    =&\sum_{\bk}
    \rho(\bk)g_{\mu\nu}(\bk)
    + 
    \mathrm{Cov}_{\rho(\bk)}
    \left[
    X_\mu(\bk),X_\nu(\bk)
    \right] 
    \\ &+
    \sum_{\bk}
    \partial_\mu\sqrt{\rho(\bk)}
    \partial_\nu\sqrt{\rho(\bk)}
    .
\end{split}
\label{eq:main_metric_decomposition}
\end{equation}
Here $g_{\mu\nu}(\bk) = \frac{1}{2}\Tr\left[\partial_\mu \mathcal{P}(\bk)\partial_\nu \mathcal{P}(\bk)\right]$ is the quantum metric, and covariance matrix is given by
$\mathrm{Cov}_{\rho}
    \left[
    X_\mu,X_\nu
    \right]
    =
    \sum_{\bk}\rho(\bk)X_\mu(\bk)X_\nu(\bk) 
    -
    \left[
    \sum_{\bk}\rho(\bk)X_\mu(\bk)
    \right]
    \left[
    \sum_{\bk}\rho(\bk)X_\nu(\bk)
    \right]$.
Since the covariance trace and the gradient term are non-negative, by taking the trace of Eq.~\eqref{eq:main_metric_decomposition}, the spatial extent of quantum geometric Kondo cloud obeys the lower bound
\begin{equation}
    \xi_{\text{K}}^2
    \ge
    \sum_{\bk}
    \rho(\bk)\Tr g(\bk).
\label{eq:metric_bound}
\end{equation}
Equations~\eqref{eq:main_metric_decomposition} and~\eqref{eq:metric_bound} reveal that the spatial extent of quantum geometric Kondo cloud is controlled by three gauge-invariant contributions: weighted quantum metric, covariance of dressed Berry connection, and the gradient of hybridization distribution. 
As anticipated, this expression is fundamentally distinct from the conventional metallic estimate $\hbar v_F/k_B T_{\text{K}}$, which is singular in the flat-band limit. Intriguingly, the lower bound in Eq.~\eqref{eq:metric_bound} makes this quantum geometric origin explicit. In contrast, an atomic-limit band with a $\bk$-independent Bloch spinor and local projected hybridization, the Wannier-like structure of active bath mode in Eq.~\eqref{eq:Phi} collapses to a localized orbital with vanishing geometric lower bound. 
The lower bound here is closely related to the gauge-invariant part of Wannier spread~\cite{marzari2012maximally}, but here $\xi^2_{K,\mu\nu}$ itself is a gauge-invariant many-body observable fixed by the ground-state wave function.

\emph{Quantum geometric Kondo cloud in Lieb lattice with tunable quantum metric.}\textemdash
We next demonstrate the quantum geometric Kondo cloud in the one-dimensional Lieb lattice under various orbital selective coupling. We consider a single impurity coupled to the sites in unit cell $\bR_i=0$ and set lattice constant $a=1$. The hybridization Hamiltonian is given by $H_V=\sum_{\alpha\sigma}\left(V_{\alpha}d_{\sigma}^{\dagger}c_{0\alpha\sigma}+V_{\alpha}^*c_{0\alpha\sigma}^{\dagger}d_{\sigma}\right)$ with $V_{\alpha} = |V_\alpha|e^{i\theta_\alpha}$ and $\alpha=A,B,C$. The band structures and tight-binding hoppings are shown in Figs.~\ref{fig:1}(a) and \ref{fig:3}(a), respectively. Here, $t$ denotes the nearest-neighbor hopping, and $\pm m$ are the onsite potentials on the $A$ and $C$ sublattices. The flat-band Bloch state is $u(k) \propto (-tq(k), m, tq(k))^T$ with $q(k) = (1+e^{ik})$ in the basis $c_{k\sigma}=(c_{kA\sigma},c_{kB\sigma},c_{kC\sigma})^T$. The corresponding flat-band hybridization factor is given by $v(k) \propto \left[tq(k)\left(V_C -V_A\right) + mV_B\right]$.
Apart from an overall phase, the relative phases of the impurity-orbital couplings \(V_\alpha\) control the impurity-bath interference, producing distinct forms of \(v(k)\) and hence distinct Kondo couplings \(J_f\).

We first consider the equal-amplitude impurity coupling $V_\alpha = |V_0|e^{i\theta_{\alpha}}$ and fix the redundant global hybridization phase by setting $\theta_B=0$. With the quantum metric and $|V_0|$ fixed, both projected hybridization strength $\mathcal{V}^2$ and Kondo-cloud length $\xi_{\text{K}}$ are ontrolled entirely by the relative impurity-orbital coupling phases $\theta_A$ and $\theta_C$, as shown in Figs.~\ref{fig:2}(a) and (b). The maximum of $\mathcal{V}^2$ occurs when $\theta_C-\theta_A=\pi$ and $\theta_C=\theta_B$ modulo $2\pi$~\cite{supple}. 
Thus, apart from an unphysical global phase, the flat-band Kondo
scale and cloud size are governed by phase-coherent
impurity-bath interference.

We further examine several orbital-selective impurity channels in the one-dimensional Lieb lattice [Fig.~\ref{fig:3}(a)]: the $B$-site channel, the constructive mixed channel, the antisymmetric $A/C$ channel, and the dark symmetric $A/C$ channel. Figure~\ref{fig:3}(b) compares the corresponding hybridization distributions $\rho(k)$ with the flat-band quantum metric $g(k)$. In the $B$-site channel, $\rho(k)$ is concentrated near $k=\pi$, where $g(k)$ is maximal, thereby producing the largest geometric contribution to the Kondo-cloud length.
Figure~\ref{fig:3}(c) shows the Kondo-cloud length $\xi_{\mathrm K}$ and its weighted quantum-metric lower bound $\xi_{\mathrm K,\mathrm{lb}}$ as functions of $m/t$. Increasing $m/t$ suppresses the quantum metric and shrinks the cloud. The flat-band projection requires $J_f\ll \Delta_{\mathrm{gap}}=m$, which defines the controlled regime indicated by the shaded region. Within this regime, especially for small $m/t$, $\xi_{\mathrm K}$ can extend over several lattice constants despite the absence of a Fermi velocity. Its magnitude is set not by a kinetic length scale, but by the overlap between the hybridization distribution $\rho(k)$ and the quantum metric $g(k)$. The dark symmetric $A/C$ channel provides the sharpest illustration of impurity-bath coherence. Although the impurity couples locally to both $A$ and $C$ orbitals, their flat-band components interfere destructively, giving $v(k)=0$. Consequently, the projected hybridization weight vanishes, the active bath mode $f^\dagger$ is absent, and no molecular Kondo singlet or Kondo cloud is formed.

Taken together, these results reveal an energy-length decoupling. The many-body gap satisfies $\Delta_{\mathrm{mol}}\propto J_f$, which depends on both the projected hybridization strength and interaction scale $U$, whereas $\xi_{\mathrm K}$ is fixed by the geometry of the active bath mode and is independent of $J_f$ at fixed $\rho(\bk)$. This energy-length decoupling is a typical characteristic of flat-band systems with nontrivial quantum geometry, where logarithmic RG flow is quenched by the absence of dispersive energy shells.

\emph{Weak dispersion and perturbative stability.}\textemdash
We finally consider an isolated nearly flat band with dispersion $\xi(\bk)$ and bandwidth $W_b$. The weak-dispersion expansion is controlled when $W_b\ll\Delta_{\rm mol}=k_BT_{\text{K}}^{\text{flat}}$. Although the exchange vertex remains factorized, the kinetic energy is no longer diagonal in the active/dark-mode basis and therefore admixes dark modes into the molecular singlet. At generic filling, this mixing generates dark particle-hole configurations with amplitudes of order $W_b/\Delta_{\rm mol}$.
For an explicit analytic form, we restrict to the minimal molecular sector with an empty dark sector, where the perturbed state remains a pure total-spin singlet~\cite{supple}
\begin{equation}
\ket{\Psi_S}
=
\frac{1}{\sqrt{2}}
\sum_{\bk}
\psi_S(\bk)
\left(
d^\dagger_\uparrow \gamma^\dagger_{\bk\downarrow}
-
d^\dagger_\downarrow \gamma^\dagger_{\bk\uparrow}
\right)
\ket{0}.
\label{eq:weak_dispersion_singlet}
\end{equation}
Here we define $a(\bk)\equiv v^*(\bk)/\mathcal{V}$. To first order, $\psi_S(\bk)=a(\bk)\left[1-\delta\xi(\bk)/\Delta_{\rm mol}\right]+O[(W_b/\Delta_{\rm mol})^2]$,
with $\delta\xi(\bk) \simeq \xi(\bk) -\sum_{\bk}\rho(\bk)\xi(\bk)$ as the energy deviation from hybridization-weighted mean. This state remains normalized through first order, since $\sum_{\bk}\rho(\bk)\delta\xi(\bk)=0$.

The corresponding real-space screening orbital is $\Phi_{S,\alpha}(\bR_i)=\Phi^{(0)}_\alpha(\bR_i)-\Phi^{(1)}_\alpha(\bR_i)/\Delta_{\rm mol}+O[(W_b/\Delta_{\rm mol})^2]$, where $\Phi^{(1)}_\alpha(\bR_i)\equiv N^{-1/2}\sum_{\bk}e^{-i\bk\cdot\bR_i}\delta\xi(\bk)a(\bk)u^*_\alpha(\bk)$. Consequently, within this minimal molecular sector, the impurity--bath spin correlation retains the form $C_{i\alpha}=-3|\Phi_{S,\alpha}(\bR_i)|^2/4$, with its leading spatial deformation controlled by $W_b/\Delta_{\rm mol}$.
This molecular regime and algebraic Kondo scale agree with numerical RG studies of narrow-band Lieb lattices~\cite{tran2018molecular}. When $W_b \sim \Delta_{\rm mol}$, perturbation theory breaks down, signaling a crossover to Fermi-surface-dominated Kondo screening.

\emph{Conclusion.}\textemdash
We have identified that, in the absence of Fermi-surface kinematics, a magnetic impurity coupled to a flat-band bath with nontrivial quantum geometry forms a coherent superposition of impurity-bath singlets as the ground state, giving rise to a quantum geometric Kondo cloud whose size is lower bounded by the hybridization-weighted quantum metric. 
When the system is weakly dispersive, the ratio \(W_b/J_f\) controls the crossover between quantum-geometric molecular screening and metallic Kondo screening, providing a direct route for numerical studies in flat-band models with tunable bandwidth.

\emph{Acknowledgments.}\textemdash
We thank Jinchao Zhao, Ying-Ming Xie and Xilin Feng for valuable and inspiring discussions. We acknowledge the support of the Ministry of Science and Technology, China, The New Cornerstone Foundation, and the Hong Kong Research Grants Council through Grants No. MOST23SC01-A, No. RFS2021-6S03, No. C6053-23G, No. AoE/P-701/20, AoE/P-604/25R, No. 16309223, No. 16311424 and No. 16300325.

\bibliography{References}

@article{kondo1964resistance,
  author  = {Kondo, Jun},
  title   = {Resistance minimum in dilute magnetic alloys},
  journal = {Prog. Theor. Phys.},
  volume  = {32},
  pages   = {37},
  doi     = {10.1143/PTP.32.37},
  year    = {1964}
}

@article{anderson1961localized,
  author  = {Anderson, Philip W.},
  title   = {Localized magnetic states in metals},
  journal = {Phys. Rev.},
  volume  = {124},
  pages   = {41},
  doi     = {10.1103/PhysRev.124.41},
  year    = {1961}
}

@article{barnes1976new,
  author  = {Barnes, S. E.},
  title   = {New method for the {Anderson} model},
  journal = {J. Phys. F: Met. Phys.},
  volume  = {6},
  pages   = {1375},
  doi     = {10.1088/0305-4608/6/7/018},
  year    = {1976}
}

@book{hewson1997kondo,
  author    = {Hewson, Alexander C.},
  title     = {The {Kondo} Problem to Heavy Fermions},
  publisher = {Cambridge University Press},
  address   = {Cambridge},
  year      = {1993}
}

@article{sorensen1996scaling,
  author  = {S{\o}rensen, Erik S. and Affleck, Ian},
  title   = {Scaling theory of the {Kondo} screening cloud},
  journal = {Phys. Rev. B},
  volume  = {53},
  pages   = {9153},
  doi     = {10.1103/PhysRevB.53.9153},
  year    = {1996}
}

@article{barzykin1996kondo,
  author  = {Barzykin, Victor and Affleck, Ian},
  title   = {The {Kondo} screening cloud: What can we learn from perturbation theory?},
  journal = {Phys. Rev. Lett.},
  volume  = {76},
  pages   = {4959},
  doi     = {10.1103/PhysRevLett.76.4959},
  year    = {1996}
}

@article{v2020observation,
  author  = {Borzenets, Ivan V. and Shim, Jeongmin and Chen, Jason C. H. and Ludwig, Arne and Wieck, Andreas D. and Tarucha, Seigo and Sim, H.-S. and Yamamoto, Michihisa},
  title   = {Observation of the {Kondo} screening cloud},
  journal = {Nature (London)},
  volume  = {579},
  pages   = {210},
  doi     = {10.1038/s41586-020-2058-6},
  year    = {2020}
}

@article{borda2007kondo,
  author  = {Borda, L{\'a}szl{\'o}},
  title   = {{Kondo} screening cloud in a one-dimensional wire: Numerical renormalization group study},
  journal = {Phys. Rev. B},
  volume  = {75},
  pages   = {041307(R)},
  doi     = {10.1103/PhysRevB.75.041307},
  year    = {2007}
}

@article{han2024quantum,
  author  = {Han, Zhaoyu and Herzog-Arbeitman, Jonah and Bernevig, B. Andrei and Kivelson, Steven A.},
  title   = {``Quantum geometric nesting'' and solvable model flat-band systems},
  journal = {Phys. Rev. X},
  volume  = {14},
  pages   = {041004},
  doi     = {10.1103/PhysRevX.14.041004},
  year    = {2024}
}

@article{zhang2026identifying,
  author  = {Zhang, Jia-Xin and Wang, Wen O. and Balents, Leon and Savary, Lucile},
  title   = {Identifying instabilities with quantum geometry in flat-band systems},
  journal = {Phys. Rev. Lett.},
  volume  = {136},
  pages   = {176504},
  doi     = {10.1103/rw6g-w7my},
  year    = {2026}
}

@article{gao2026bootstrapping,
  author  = {Gao, Qiang and Han, Zhaoyu and Khalaf, Eslam},
  title   = {Bootstrapping flatband superconductors: Rigorous lower bounds on superfluid stiffness},
  journal = {Phys. Rev. Lett.},
  volume  = {136},
  pages   = {076503},
  doi     = {10.1103/gw85-5r92},
  year    = {2026}
}

@article{chen2024ginzburg,
  author  = {Chen, Shuai A. and Law, K. T.},
  title   = {{Ginzburg--Landau} theory of flat-band superconductors with quantum metric},
  journal = {Phys. Rev. Lett.},
  volume  = {132},
  pages   = {026002},
  doi     = {10.1103/PhysRevLett.132.026002},
  year    = {2024}
}

@article{gubernatis1987spin,
  author  = {Gubernatis, J. E. and Hirsch, J. E. and Scalapino, D. J.},
  title   = {Spin and charge correlations around an {Anderson} magnetic impurity},
  journal = {Phys. Rev. B},
  volume  = {35},
  pages   = {8478},
  doi     = {10.1103/PhysRevB.35.8478},
  year    = {1987}
}

@article{affleck2001detecting,
  author  = {Affleck, Ian and Simon, Pascal},
  title   = {Detecting the {Kondo} screening cloud around a quantum dot},
  journal = {Phys. Rev. Lett.},
  volume  = {86},
  pages   = {2854},
  doi     = {10.1103/PhysRevLett.86.2854},
  year    = {2001}
}

@article{sorensen2005kondo,
  author  = {S{\o}rensen, Erik S. and Affleck, Ian},
  title   = {{Kondo} screening cloud around a quantum dot: Large-scale numerical results},
  journal = {Phys. Rev. Lett.},
  volume  = {94},
  pages   = {086601},
  doi     = {10.1103/PhysRevLett.94.086601},
  year    = {2005}
}

@article{hand2006spin,
  author  = {Hand, Thomas and Kroha, Johann and Monien, Hartmut},
  title   = {Spin correlations and finite-size effects in the one-dimensional {Kondo} box},
  journal = {Phys. Rev. Lett.},
  volume  = {97},
  pages   = {136604},
  doi     = {10.1103/PhysRevLett.97.136604},
  year    = {2006}
}

@article{affleck2008friedel,
  author  = {Affleck, Ian and Borda, L{\'a}szl{\'o} and Saleur, Hubert},
  title   = {Friedel oscillations and the {Kondo} screening cloud},
  journal = {Phys. Rev. B},
  volume  = {77},
  pages   = {180404(R)},
  doi     = {10.1103/PhysRevB.77.180404},
  year    = {2008}
}

@article{pereira2008kondo,
  author  = {Pereira, Rodrigo G. and Laflorencie, Nicolas and Affleck, Ian and Halperin, Bertrand I.},
  title   = {{Kondo} screening cloud and charge staircase in one-dimensional mesoscopic devices},
  journal = {Phys. Rev. B},
  volume  = {77},
  pages   = {125327},
  doi     = {10.1103/PhysRevB.77.125327},
  year    = {2008}
}

@article{simon2020contrasting,
  author  = {Simon, Steven H. and Rudner, Mark S.},
  title   = {Contrasting lattice geometry dependent versus independent quantities: Ramifications for {Berry} curvature, energy gaps, and dynamics},
  journal = {Phys. Rev. B},
  volume  = {102},
  pages   = {165148},
  doi     = {10.1103/PhysRevB.102.165148},
  year    = {2020}
}

@article{marzari2012maximally,
  author  = {Marzari, Nicola and Mostofi, Arash A. and Yates, Jonathan R. and Souza, Ivo and Vanderbilt, David},
  title   = {Maximally localized {Wannier} functions: Theory and applications},
  journal = {Rev. Mod. Phys.},
  volume  = {84},
  pages   = {1419},
  doi     = {10.1103/RevModPhys.84.1419},
  year    = {2012}
}

@article{anderson1970poor,
  author  = {Anderson, P. W.},
  title   = {A poor man's derivation of scaling laws for the {Kondo} problem},
  journal = {J. Phys. C: Solid State Phys.},
  volume  = {3},
  pages   = {2436},
  doi     = {10.1088/0022-3719/3/12/008},
  year    = {1970}
}

@article{hu2019geometric,
  author  = {Hu, Xiang and Hyart, Timo and Pikulin, Dmitry I. and Rossi, Enrico},
  title   = {Geometric and conventional contribution to the superfluid weight in twisted bilayer graphene},
  journal = {Phys. Rev. Lett.},
  volume  = {123},
  pages   = {237002},
  doi     = {10.1103/PhysRevLett.123.237002},
  year    = {2019}
}

@article{peotta2015superfluidity,
  author  = {Peotta, Sebastiano and T{\"o}rm{\"a}, P{\"a}ivi},
  title   = {Superfluidity in topologically nontrivial flat bands},
  journal = {Nat. Commun.},
  volume  = {6},
  pages   = {8944},
  doi     = {10.1038/ncomms9944},
  year    = {2015}
}

@article{liang2017band,
  author  = {Liang, Long and Vanhala, Tuomas I. and Peotta, Sebastiano and Siro, Topi and Harju, Ari and T{\"o}rm{\"a}, P{\"a}ivi},
  title   = {Band geometry, {Berry} curvature, and superfluid weight},
  journal = {Phys. Rev. B},
  volume  = {95},
  pages   = {024515},
  doi     = {10.1103/PhysRevB.95.024515},
  year    = {2017}
}

@article{herzog2022superfluid,
  author  = {Herzog-Arbeitman, Jonah and Peri, Valerio and Schindler, Frank and Huber, Sebastian D. and Bernevig, B. Andrei},
  title   = {Superfluid weight bounds from symmetry and quantum geometry in flat bands},
  journal = {Phys. Rev. Lett.},
  volume  = {128},
  pages   = {087002},
  doi     = {10.1103/PhysRevLett.128.087002},
  year    = {2022}
}

@article{huhtinen2022revisiting,
  author  = {Huhtinen, Kukka-Emilia and Herzog-Arbeitman, Jonah and Chew, Aaron and Bernevig, B. Andrei and T{\"o}rm{\"a}, P{\"a}ivi},
  title   = {Revisiting flat band superconductivity: Dependence on minimal quantum metric and band touchings},
  journal = {Phys. Rev. B},
  volume  = {106},
  pages   = {014518},
  doi     = {10.1103/PhysRevB.106.014518},
  year    = {2022}
}

@article{hofmann2023superconductivity,
  author  = {Hofmann, Johannes S. and Berg, Erez and Chowdhury, Debanjan},
  title   = {Superconductivity, charge density wave, and supersolidity in flat bands with a tunable quantum metric},
  journal = {Phys. Rev. Lett.},
  volume  = {130},
  pages   = {226001},
  doi     = {10.1103/PhysRevLett.130.226001},
  year    = {2023}
}

@article{hu2025anomalous,
  author  = {Hu, Jin-Xin and Chen, Shuai A. and Law, K. T.},
  title   = {Anomalous coherence length in superconductors with quantum metric},
  journal = {Commun. Phys.},
  volume  = {8},
  pages   = {20},
  doi     = {10.1038/s42005-024-01930-0},
  year    = {2025}
}

@article{sun2025flat,
  author  = {Sun, Zi-Ting and Yu, Ruo-Peng and Chen, Shuai A. and Hu, Jin-Xin and Law, K. T.},
  title   = {Flat-band {Fulde--Ferrell--Larkin--Ovchinnikov} state from quantum geometric discrepancy},
  journal = {Quantum Front.},
  volume  = {4},
  pages   = {20},
  doi     = {10.1007/s44214-025-00093-5},
  year    = {2025}
}

@article{cao2018unconventional,
  author  = {Cao, Yuan and Fatemi, Valla and Fang, Shiang and Watanabe, Kenji and Taniguchi, Takashi and Kaxiras, Efthimios and Jarillo-Herrero, Pablo},
  title   = {Unconventional superconductivity in magic-angle graphene superlattices},
  journal = {Nature (London)},
  volume  = {556},
  pages   = {43},
  doi     = {10.1038/nature26160},
  year    = {2018}
}

@article{cao2018correlated,
  author  = {Cao, Yuan and Fatemi, Valla and Demir, Ahmet and Fang, Shiang and Tomarken, Spencer L. and Luo, Jason Y. and Sanchez-Yamagishi, Javier D. and Watanabe, Kenji and Taniguchi, Takashi and Kaxiras, Efthimios and others},
  title   = {Correlated insulator behaviour at half-filling in magic-angle graphene superlattices},
  journal = {Nature (London)},
  volume  = {556},
  pages   = {80},
  doi     = {10.1038/nature26154},
  year    = {2018}
}

@article{bistritzer2011moire,
  author  = {Bistritzer, Rafi and MacDonald, Allan H.},
  title   = {Moir{\'e} bands in twisted double-layer graphene},
  journal = {Proc. Natl. Acad. Sci. U.S.A.},
  volume  = {108},
  pages   = {12233},
  doi     = {10.1073/pnas.1108174108},
  year    = {2011}
}

@article{po2018origin,
  author  = {Po, Hoi Chun and Zou, Liujun and Vishwanath, Ashvin and Senthil, T.},
  title   = {Origin of {Mott} insulating behavior and superconductivity in twisted bilayer graphene},
  journal = {Phys. Rev. X},
  volume  = {8},
  pages   = {031089},
  doi     = {10.1103/PhysRevX.8.031089},
  year    = {2018}
}

@article{song2022magic,
  author  = {Song, Zhi-Da and Bernevig, B. Andrei},
  title   = {Magic-angle twisted bilayer graphene as a topological heavy fermion problem},
  journal = {Phys. Rev. Lett.},
  volume  = {129},
  pages   = {047601},
  doi     = {10.1103/PhysRevLett.129.047601},
  year    = {2022}
}

@article{li2025flat,
  author  = {Li, Zhong C. F. and Deng, Yuxuan and Chen, Shuai A. and Efetov, Dmitri K. and Law, K. T.},
  title   = {Flat band {Josephson} junctions with quantum metric},
  journal = {Phys. Rev. Research},
  volume  = {7},
  pages   = {023273},
  doi     = {10.1103/PhysRevResearch.7.023273},
  year    = {2025}
}

@misc{dai2026quantummetriclocalizationquantum,
  author        = {Dai, Wen-Bo and Zhao, Jinchao and Chen, Shuai A. and Law, K. T.},
  title         = {Quantum Metric Localization and Quantum Metric Protection},
  eprint        = {2605.03987},
  archiveprefix = {arXiv},
  primaryclass  = {cond-mat.mes-hall},
  year          = {2026}
}

@misc{chau2026quantum,
  author        = {Chau, Chun Wang and Xiang, Tian and Chen, Shuai A. and Law, K. T.},
  title         = {Quantum Metric Length as a Fundamental Length Scale in Disordered Flat Band Materials},
  eprint        = {2602.01354},
  archiveprefix = {arXiv},
  primaryclass  = {cond-mat.mes-hall},
  year          = {2026}
}

@article{manoharan2000quantum,
  author  = {Manoharan, H. C. and Lutz, C. P. and Eigler, D. M.},
  title   = {Quantum mirages formed by coherent projection of electronic structure},
  journal = {Nature (London)},
  volume  = {403},
  pages   = {512},
  doi     = {10.1038/35000508},
  year    = {2000}
}

@article{van2024modulated,
  author  = {van Efferen, Camiel and Fischer, Jeison and Costi, Theo A. and Rosch, Achim and Michely, Thomas and Jolie, Wouter},
  title   = {Modulated {Kondo} screening along magnetic mirror twin boundaries in monolayer {MoS}$_2$},
  journal = {Nat. Phys.},
  volume  = {20},
  pages   = {82},
  doi     = {10.1038/s41567-023-02250-w},
  year    = {2024}
}

@article{knorr2002kondo,
  author  = {Knorr, Nikolaus and Schneider, M. Alexander and Diekh{\"o}ner, Lars and Wahl, Peter and Kern, Klaus},
  title   = {{Kondo} effect of single {Co} adatoms on {Cu} surfaces},
  journal = {Phys. Rev. Lett.},
  volume  = {88},
  pages   = {096804},
  doi     = {10.1103/PhysRevLett.88.096804},
  year    = {2002}
}

@article{zhao2023gate,
  author  = {Zhao, Wenjin and Shen, Bowen and Tao, Zui and Han, Zhongdong and Kang, Kaifei and Watanabe, Kenji and Taniguchi, Takashi and Mak, Kin Fai and Shan, Jie},
  title   = {Gate-tunable heavy fermions in a moir{\'e} {Kondo} lattice},
  journal = {Nature (London)},
  volume  = {616},
  pages   = {61},
  doi     = {10.1038/s41586-023-05800-7},
  year    = {2023}
}

@article{zhao2024emergence,
  author  = {Zhao, Wenjin and Shen, Bowen and Tao, Zui and Kim, Sunghoon and Kn{\"u}ppel, Patrick and Han, Zhongdong and Zhang, Yichi and Watanabe, Kenji and Taniguchi, Takashi and Chowdhury, Debanjan and others},
  title   = {Emergence of ferromagnetism at the onset of moir{\'e} {Kondo} breakdown},
  journal = {Nat. Phys.},
  volume  = {20},
  pages   = {1772},
  doi     = {10.1038/s41567-024-02636-4},
  year    = {2024}
}

@misc{supple,
  key  = {Supplemental Material},
  note = {see Supplemental Material for technical details of the Schrieffer--Wolff transformation, the active-mode solution, the Kondo-cloud correlation and cloud-size tensor, the phase dependence of the projected hybridization, and the weak-dispersion expansion.}
}

@article{bulla2008numerical,
  author  = {Bulla, Ralf and Costi, Theo A. and Pruschke, Thomas},
  title   = {Numerical renormalization group method for quantum impurity systems},
  journal = {Rev. Mod. Phys.},
  volume  = {80},
  pages   = {395},
  doi     = {10.1103/RevModPhys.80.395},
  year    = {2008}
}

@article{tran2018molecular,
  author  = {Tran, Minh-Tien and Nguyen, Thuy Thi},
  title   = {Molecular {Kondo} effect in flat-band lattices},
  journal = {Phys. Rev. B},
  volume  = {97},
  pages   = {155125},
  doi     = {10.1103/PhysRevB.97.155125},
  year    = {2018}
}

\end{document}